\documentclass[12pt]{article}
\pdfoutput=1
\usepackage{putex}
\usepackage{feyn}
\usepackage[vcentermath]{youngtab}

\usepackage{graphicx}
\usepackage{epstopdf}
\usepackage{enumerate}
\usepackage{cite}
\usepackage{tensor}
\usepackage{slashed}
\usepackage{feynmf}

\usepackage{hyperref}

\numberwithin{equation}{section}

\newcommand {\be} {\begin {equation}}
\newcommand {\ee} {\end {equation}}

\newcommand {\bes} {\begin {equation*}}
\newcommand {\ees} {\end {equation*}}

\newcommand{\es}[2] {\begin{equation} \label{#1} \begin{split} #2 \end{split} \end{equation}}

\newcommand{\beq}{\begin{equation}}
\newcommand{\eeq}{\end{equation}}

\def\be{ \begin{equation} }
\def\ee{ \end{equation} }

\begin{document}

\preprint{PUPT-2458}

\institution{PU}{Department of Physics, Princeton University, Princeton, NJ 08544}
\institution{PCTS}{Princeton Center for Theoretical Science, Princeton University, Princeton, NJ 08544}

\title{
Higher Spin AdS$_{d+1}$/CFT$_d$ at One Loop
}

\authors{Simone Giombi,\worksat{\PU} Igor R.~Klebanov\worksat{\PU,\PCTS} and Benjamin R.~Safdi\worksat{\PU}
}

\abstract{Following \cite{Giombi:2013fka}, we carry out one loop tests of higher spin AdS$_{d+1}$/CFT$_d$ correspondences for $d\geq 2$.
The Vasiliev theories in AdS$_{d+1}$, which contain each integer spin once, are related to the $U(N)$ singlet sector of the $d$-dimensional CFT of $N$ free complex scalar fields;
the minimal theories containing each even spin once -- to the $O(N)$ singlet sector of the CFT of $N$ free real scalar fields.
Using analytic continuation of higher spin zeta functions, which naturally regulate the spin sums,
we calculate one loop vacuum energies in Euclidean AdS$_{d+1}$.
In even $d$ we compare the result with the $O(N^0)$ correction to the $a$-coefficient of the Weyl anomaly; in odd $d$ -- with the $O(N^0)$ correction to the free energy $F$ on the $d$-dimensional sphere.
For the theories of integer spins, the correction vanishes in agreement with the CFT of $N$ free complex scalars. For the minimal theories, the correction
always equals the contribution of one real conformal scalar field in $d$ dimensions. As explained in \cite{Giombi:2013fka},
this result may agree with the
$O(N)$ singlet sector of the theory of $N$ real scalar fields, provided the coupling constant in the higher spin theory is identified as $G_N\sim 1/(N-1)$.
Our calculations in even $d$ are closely related to finding the regularized $a$-anomalies of conformal higher spin theories.
In each even $d$ we identify two such theories with vanishing $a$-anomaly: a theory of all integer spins, and a theory of all even spins coupled to a complex conformal scalar.
We also discuss an interacting UV fixed point in $d=5$ obtained from the free scalar theory via an irrelevant double-trace quartic interaction.
This interacting large $N$ theory is dual to the Vasiliev theory in AdS$_6$ where the bulk scalar is quantized with the alternate
boundary condition.
}

\date{}
\maketitle

\tableofcontents

\section{Introduction and Summary}

The correspondence between large $N$ supersymmetric gauge theory and string theory in AdS backgrounds \cite{Maldacena:1997re,Gubser:1998bc,Witten:1998qj} has been a fruitful area of research for quite some time.
Aside from teaching us valuable lessons about quantum gravity, the AdS/CFT correspondence leads to many concrete predictions about the behavior of strongly coupled gauge theory. While a great deal of evidence has been accumulated in favor of the
correspondence (see \cite{Beisert:2010jr} for a review of the results found using exact integrability), a proof is yet to be found. This is probably due to the fact that the AdS/CFT correspondence often relates pairs of very complicated theories, for example
the maximally supersymmetric $SU(N)$ gauge theory in four dimensions and type IIB string theory on AdS$_5\times$ S$^5$.

In this paper we will explore a different class of AdS/CFT conjectures, where the dynamical fields in the CFT are $N$-component vectors rather than $N\times N$ matrices. This makes such ``vectorial" large $N$ field theories
more
tractable, and a number of concrete AdS/CFT conjectures for them have been made
\cite{Klebanov:2002ja,Sezgin:2003pt,Leigh:2003gk,Giombi:2011kc, Aharony:2011jz, Chang:2012kt,Gaberdiel:2010pz,Gaberdiel:2012uj,Chang:2011mz} (see \cite{Giombi:2012ms} for a review).
For instance, consider a $d$-dimensional CFT of $N$ free complex or real scalar fields restricted to the $U(N)$ or $O(N)$ singlet
sector \cite{Klebanov:2002ja}. Such theories possess an infinite set of higher-spin conserved currents; therefore, the dual
AdS$_{d+1}$ description must contain an infinite set of interacting massless gauge fields \cite{Sundborg:2000wp}. Thanks to the many years of work by Vasiliev and collaborators, the classical
non-linear equations for such theories have been found in various dimensions \cite{Fradkin:1987ks,Vasiliev:1990en,Vasiliev:1992av,Vasiliev:1995dn,Prokushkin:1998bq,Vasiliev:1999ba, Vasiliev:2003ev, Bekaert:2005vh}. In particular, a formulation of the higher spin theory in AdS$_{d+1}$ for arbitrary dimension $d$ was obtained in \cite{Vasiliev:2003ev}. The spectrum of the simplest version of this model contains a tower of totally symmetric massless gauge fields, one for each positive integer spin, plus a scalar with $m^2=-2(d-2)/\ell^2$ (here $\ell$ is the AdS radius of curvature,
which we will often set to 1). This value of $m^2$ is above the Breitenlohner-Freedman \cite{Breitenlohner:1982jf}
bound $m^2_{BF}=-d^2 / (4 \ell^2)$ in any dimension except $d=4$, where it is right at the bound. According to the AdS/CFT dictionary, a bulk scalar with $m^2=-2(d-2)/\ell^2$ is dual to a scalar primary with dimension $\Delta=d-2$ or $\Delta=2$ \cite{Klebanov:1999tb}. The former value of $\Delta$ is the dimension of the operator $\bar \phi^i\phi^i$ in the free scalar theory.\footnote{The other possible dimension, $\Delta=2$, is above the unitarity bound only for $d<6$. For $2<d<4$, it corresponds to the dimension of the scalar operator in the Wilson-Fisher IR fixed point of the large $N$ theory with $\frac {\lambda} {4} (\bar \phi^i\phi^i)^2$ interaction. This critical theory in $d=3$ corresponds to Vasiliev's type A theory in AdS$_4$ with the $\Delta=2$ scalar boundary condition \cite{Klebanov:2002ja}. For $d>4$,  there exists a large $N$ UV fixed point of the interacting scalar theory at a negative value of $\lambda$ \cite{Maldacena:2012sf}. This UV fixed point in $d=5$, which should correspond to Vasiliev's theory in AdS$_6$ with the $\Delta=2$ scalar boundary condition, will be discussed in Section \ref{5dfixed}.} The standard boundary conditions on the massless spin $s$ gauge fields put them in correspondence with the conserved currents of dimension
$\Delta=d-2+s$. Thus, the spectrum of this class of Vasiliev theories \cite{Vasiliev:2003ev} is in one-to-one correspondence with the single trace primary operators
of the free $d$-dimensional complex scalar CFT restricted to the $U(N)$ singlet sector. One may also impose a consistent truncation of the bulk theory that retains the even spins only; such a minimal Vasiliev theory is conjectured to be dual to the $O(N)$ singlet sector of the $d$-dimensional CFT of $N$ real scalars. In $d=3$ a well-known way of imposing the singlet constraint is by coupling to the Chern-Simons gauge theory \cite{Giombi:2011kc, Aharony:2011jz}; in higher $d$, one could contemplate coupling to some topological gauge theory.

In AdS$_4$, the Vasiliev non-linear equations have been formulated in a particularly nice form by using twistor-like variables \cite{Vasiliev:1990en,Vasiliev:1992av}. There exist two parity invariant versions of the AdS$_4$ Vasiliev theory, commonly called type A and B. In the former the spin zero field has positive parity, while in the latter it has negative parity. The type A Vasiliev theory, which includes each integer spin once, has been conjectured to be dual to the $U(N)$ singlet sector of the theory of $N$ massless complex scalar fields \cite{Klebanov:2002ja}, while the type B theory has been conjectured to be dual to the theory of $N$ massless fermions \cite{Sezgin:2003pt, Leigh:2003gk}.  There also exists a projection of the type A/B theory to the minimal theory containing even spins only, which has been conjectured to be dual to the $O(N)$ singlet sector of the theory of $N$ real massless scalars/fermions. At the level of the 3-point functions these conjectures have been checked at leading order in $N$
\cite{Giombi:2009wh,Giombi:2010vg,Giombi:2011ya,Maldacena:2011jn,Maldacena:2012sf,Colombo:2012jx,Didenko:2012tv,Gelfond:2013xt,Didenko:2013bj}.
The Vasiliev theory in general dimensions \cite{Vasiliev:2003ev}, whose spectrum is dual to that of the $U(N)$ or $O(N)$ singlet sector of
$d$-dimensional scalar CFT, is a generalization of the type A model in AdS$_4$.\footnote{The higher spin theory dual to massless free fermions in general dimensions would have to involve higher spin fields of more general mixed symmetry type \cite{Vasiliev:2004cm}, and the corresponding non-linear theory has not been constructed in detail yet.} A first step towards computing holographic correlation functions in such higher spin theory in general dimensions has been taken in \cite{Didenko:2012vh}.

Another possible test of the higher-spin AdS/CFT dualities, suggested in \cite{Klebanov:2011gs},
is to match the $d$-sphere free energy $F=-\log Z_{S^d}$ with a corresponding
calculation of integrated vacuum energy in Euclidean AdS$_{d+1}$, i.e. the hyperbolic space H$_{d+1}$. The bulk partition function takes the form
\begin{equation}
Z_{\rm bulk} = e^{-\frac{1}{G_N} F^{(0)}-F^{(1)}-G_N F^{(2)}+\ldots} = e^{-F_{\rm bulk}}\,.
\label{Zbulk}
\end{equation}
The leading classical term of order $1/G_N$ in the bulk must match the term of order $N$ in the field theory free energy on $S^d$. Unfortunately, such a comparison appears to be hard, since the classical action for Vasiliev theory is currently not well understood.
Luckily, it is known that the fluctuations of each massless spin $s$ field must be governed by the standard quadratic actions in AdS$_{d+1}$, and the one-loop correction to the vacuum energy is known for each
spin \cite{Camporesi:1993mz,Camporesi:1994ga}.
In \cite{Giombi:2013fka} a regularized sum over these vacuum energies, $F^{(1)}$, was evaluated in H$_4$ and compared with the $O(N^0)$ term in the 3-sphere
free energy. For the Vasiliev theory including fields of all positive integer spin $s$, corresponding ghosts of spin $s-1$,
and a scalar with the boundary condition corresponding to operator dimension $\Delta=1$, the regularized
sum was found to vanish: $F^{(1)}=0$. This is the expected result for the $O(N^0)$ correction to $F$ in the theory of $N$ free complex scalar fields.
A more surprising result was found for the minimal Vasiliev theory with even spins only, where the regularized sum turned out to be \cite{Giombi:2013fka}\footnote{Closely related results were later found \cite{Jevicki:2014mfa} using the collective field approach to the $d=3$ scalar theories \cite{Das:2003vw, Koch:2010cy, Jevicki:2012fh}, where it is important to include the contributions of the path integral measure.}
\begin{equation}
F^{(1)}_{{\rm min}} = \frac{\log 2}{8}-\frac{3\zeta(3)}{16\pi^2}\, .
\end{equation}
This equals the $F$-value for a single real scalar field \cite{Klebanov:2011gs}. This one loop correction may be consistent with the duality proposed in
\cite{Klebanov:2002ja}, provided there
is a shift $N\rightarrow N-1$ in the identification of the bulk coupling constant: $1/G_N \sim N-1$.

A similar sum over one loop vacuum energies was also briefly discussed for the Vasiliev theories in Euclidean AdS$_5$ \cite{Giombi:2013fka}. In this case, the regularized volume of EAdS$_5$, i.e. the hyperbolic space H$_5$, contains
a logarithmic divergence $\sim \log R$, where $R$ is the radius of $S^4$ at the boundary. Hence, the appropriately normalized vacuum energy must be compared with the
$a$-coefficient of the Weyl anomaly in the dual field theory. It was found \cite{Giombi:2013yva,Giombi:2013fka} that the contribution of the $s=0$ field vanishes, while for each $s>0$ the appropriately gauge fixed contribution is
\begin{equation}
\label{acontr}
-\frac{1}{360} s^2 (1+s)^2 [3+ 14 s (1+s)]\ .
\end{equation}
Using the Riemann zeta-function to regularize the divergent sum over $s$ \cite{Giombi:2013yva,Giombi:2013fka}, the sum over all integer spins vanishes:
$a^{(1)}=0$. In the minimal theory the sum over even spins is \cite{Giombi:2013fka}
\begin{equation}
a^{(1)}_{\rm min}= -\frac{1}{45} \left ( 20 \zeta (-3) + 168 \zeta(-5) \right )=\frac {1} {90}\ ,
\end{equation}
which is the $a$-coefficient for a real scalar field in $d=4$.

Interestingly, (\ref{acontr}) is $-\frac 1 2$ times the anomaly $a$-coefficient of the Fradkin-Tseytlin conformal spin $s$ theory
in four dimensions \cite{Fradkin:1985am}, recently calculated for general spin in
\cite{Giombi:2013yva, Tseytlin:2013jya}. Thus, the vanishing of the regularized sum of (\ref{acontr}) over all integer spins is related to the fact that the
the $d=4$ interacting conformal higher spin theory, which includes each integer spin once \cite{Tseytlin:2002gz,Segal:2002gd},
has a vanishing Weyl $a$-anomaly \cite{Giombi:2013yva}. The $d=4$ interacting conformal higher spin theory including each even spin once has the zeta function regularized Weyl anomaly $a$-coefficient $-1/45$, which is minus that of a conformal complex scalar field.
Thus, it is plausible that there exists a consistent $d=4$ conformal theory of even spin fields coupled to a conformal complex scalar field.

In this paper we will show that the results found in \cite{Giombi:2013fka} for Euclidean AdS$_4$ and AdS$_5$ generalize to all AdS$_{d+1}$ with $d\geq 2$.
Namely, we will show that the appropriately regularized sum over one loop vacuum energies vanishes for the theory including all integer spins and equals the contribution of a single real conformal scalar field on $S^d$ for the minimal theory with even spins only.\footnote{
In \cite{Giombi:2013fka} it was also conjectured that the $USp (N)$ singlet sector of the theory of $N$ free complex scalar
fields, where $N$ is even, is dual to the $husp(2;0|4)$ Vasiliev theory in AdS$_4$. The latter theory contains one field of each even spin and three
fields of each odd spin \cite{Vasiliev:1999ba}. This conjecture can also be generalized to all $d\geq 2$ and tested using the same calculations as are necessary
for the integer spin and the even spin theories. For all $d$ the regularized sum over one loop vacuum energies in AdS$_{d+1}$ is then found to be equal to minus the contribution of a conformal complex field on $S^d$. Taking $G_N^{-1}\sim N+1$ for this theory can provide consistency with the conjectured $USp(N)$ AdS$_{d+1}$/CFT$_d$ duality.}
 For a conformal theory in  odd $d$, the $S^d$ free energy is a well-defined
finite number independent of the sphere radius $R$, which is obtained after subtracting power-law divergences. In even $d$, one instead has $F= a \log R$, where
$a$ is a Weyl anomaly coefficient. With this normalization, the Weyl anomaly of a real conformal scalar in $d$ dimensions is $a_{S}=-1/3, 1/90, -1/756, 23/113400$
in $d=2,4,6,8$ respectively (for different derivations of these numbers see, for example, \cite{Casini:2010kt} or eq. (9.2) of \cite{Giombi:2013yva}).
Our calculations in even $d$ amount to AdS/CFT matching of the Weyl anomaly $a$-coefficients at $O(N^0)$.
They are also closely related \cite{Giombi:2013yva,Giombi:2013fka} to finding the $a$-anomalies of conformal higher spin (CHS) theories:
 $a^{\rm CHS}= - 2 a^{(1)}$. In each even $d$ we, therefore, find two
candidate CHS theories with vanishing $a$-anomaly: a theory of all integer spins, and a theory of all even spins coupled to a standard conformal complex scalar
(these CHS theories also include a spin zero field of dimension 2 with a quadratic action that contains $d-4$ derivatives; it contributes to the Weyl anomaly in $d\neq 4$).

In evaluating the spin sums in $d>4$ we will find that the standard Riemann zeta-function regularization is no longer appropriate. A procedure to compute the regularized functional determinants that appears to be well-defined and unambiguous is to calculate the spectral zeta-function $\zeta(z)$ \cite{Hawking:1976ja} and analytically continue in
the spectral parameter $z$. Camporesi and Higuchi \cite{Camporesi:1993mz,Camporesi:1994ga} have found compact expressions for the spectral zeta function $\zeta_{(\Delta,s)}(z)$ of a field of given spin corresponding to an operator of dimension $\Delta$ in the boundary theory. The analytic continuation in the spectral parameter $z$ nicely regulates the sum over eigenvalues for each field. Luckily, for a sufficiently large $z$, the infinite sum over the tower of higher spin fields converges as well.
Therefore, we will define the complete ``higher spin zeta function" via
\es{HSDeff}{
\zeta_\text{HS}(z) = \zeta_{(d-2,0)}(z) + \sum_{s >0 } \big( \zeta_{(s+d-2,s)}(z) -  \zeta_{(s+d-1,s-1)}(z) \big) \, ,
}
where the second term under the sum subtracts the contributions of the spin $s-1$ ghost fields. In other words, we use the spectral parameter $z$ to regulate both the sum over eigenvalues for each field and the sum over the infinite tower of fields. Then, the bulk calculation is neatly summarized by
\es{HSZ}{
F^{(1)} =  - {1 \over 2} \zeta'_\text{HS}(0) - \zeta_\text{HS}(0) \log( \ell\, \Lambda) \, ,
}
where
$\Lambda$ is the UV cutoff. The logarithmic term arises in odd $d$, namely even dimensional bulk space-time, while it vanishes identically in even $d$. For classically conformal fields (such as the scalar or the $s=1$ field in the AdS$_4$ theory), it is related to the bulk conformal anomaly. Note that, unless $\zeta_\text{HS}(0)=0$, the finite part of $F^{(1)}$ is ambiguous because it can be changed by redefining the cutoff. As a simple but non-trivial consistency check, we will show that indeed $\zeta_\text{HS}(0)=0$ in all $d$, so that the Vasiliev theory \cite{Vasiliev:2003ev} is free of the logarithmic divergence in any dimension.

In even dimensions $d$ we are able to find compact expressions for the $\zeta_\text{HS}(z)$, while in odd $d$ the calculations are more complicated.
In all dimensions, we find that the regularization via the analytic continuation of $\zeta_\text{HS}(z)$ is equivalent to a simpler procedure where $z$
is continued to zero for each spin separately and then the sum over spins is evaluated using an appropriately shifted Riemann-Hurwitz zeta function, i.e.
\es{altReg: 2}{
\zeta_\text{HS}(0) = &\zeta_{(d - 2, 0)}(0) + \lim_{\alpha \to 0} \sum_{s > 0}\left(s +  {d - 3 \over 2} \right)^{- \alpha} \left( \zeta_{(d + s - 2, s)}(0) - \zeta_{(d + s - 1, s-1)}(0) \right) \, ,  \\
\zeta'_\text{HS}(0) = &\zeta'_{(d - 2, 0)}(0) + \lim_{\alpha \to 0} \sum_{s > 0}\left(s +  {d - 3 \over 2} \right)^{- \alpha} \left( \zeta'_{(d + s - 2, s)}(0) - \zeta'_{(d + s - 1, s-1)}(0) \right) \, .
}
In the next section we review the results of Camporesi and Higuchi \cite{Camporesi:1993mz,Camporesi:1994ga} and provide more detail about the calculational set-up in general dimensions.

\section{The higher-spin spectral zeta function}

Consider the Vasiliev theory in $\text{AdS}_{d+1}$ \cite{Vasiliev:2003ev}; its field content is a scalar of mass $m^2=-2(d-2)$ and a tower of totally symmetric higher spin gauge fields of all integer spins. Upon gauge fixing the linearized gauge invariance of the quadratic action, the contribution to the one-loop partition function of each massless field of spin $s$ is given by the ratio of determinants \cite{Gaberdiel:2010xv, Gaberdiel:2010ar, Gupta:2012he}
\begin{equation}
Z_s = \frac{\left[{\rm det}^{STT}_{s-1}\left(-\nabla^2+(s+d-2)(s-1)\right)\right]^{\frac{1}{2}}}{\left[{\rm det}^{STT}_{s}\left(-\nabla^2+(s+d-2)(s-2)-s\right)\right]^{\frac{1}{2}}}\ ,
\label{Zs}
\end{equation}
where the determinants are taken in the space of symmetric traceless transverse (STT) fields. Essentially, the spin $s$ contribution arises from the gauge field while the spin $s-1$ contribution is from the ghost.

These determinants can be computed by heat kernel techniques, or equivalently by means of the spectral zeta-function, which is related to the heat kernel by a Mellin transform. For a differential operator with a set of discrete eigenvalues $\lambda_n$, one usually defines the spectral zeta-function as
\begin{equation}
\zeta(z)=\sum_{n} d_n \lambda_n^{-z} \,,
\end{equation}
where $d_n$ is the degeneracy of the eigenvalue $\lambda_n$. In a non-compact space such as AdS, the sum turns into an integral over a continuous parameter $u$ labeling the eigenvalues and the degeneracy into a ``spectral density" function $\mu(u)$.\footnote{For instance, in the simple example of a scalar field on flat $\mathbb{R}^D$ with standard kinetic operator $-\nabla^2+m^2$, the eigenvalues are $\lambda(p)=p^2+m^2$, where $p$ is the momentum, the spectral density is $\mu(p)\sim p^{D-1}$, and the zeta function (per unit volume) is $\zeta(z)\sim \int_0^{\infty} dp\, p^{D-1} \left(p^2+m^2\right)^{-z}$.}

The spectral zeta-function for STT fields in $AdS_{d+1}$ with kinetic operator $-\nabla^2+\kappa^2$, where $\kappa^2$ is a constant, is known explicitly for arbitrary spin and arbitrary dimension \cite{Camporesi:1993mz,Camporesi:1994ga}. It is given by the integral
\es{spectralZeta}{
&\zeta_{(\Delta,s)}(z) = \frac{{\rm vol}(\text{AdS}_{d+1})}{{\rm vol}(S^d)}{2^{d-1} \over \pi} g(s) \int_0^\infty d u \, {\mu_s(u) \over \left[ u^2 + \nu^2 \right]^z } \,, \\
&\nu \equiv  \Delta - \frac{d}{2} \,, \qquad \Delta(\Delta-d)-s = \kappa^2 \,,
}
with $\mu_s(u)$ the spin $s$ spectral density~\cite{Camporesi:1994ga}. In even dimensions $d$
\es{spectralMuE}{
\mu_s(u) =  { \pi \left[u^2 + \left(s + {d - 2 \over 2} \right)^2 \right] \over \left[ 2^{d-1} \Gamma \left( { d+ 1 \over 2} \right) \right]^2} \prod_{j = 0}^{(d-4)/2} (u^2 + j^2)  \,,
}
while in odd $d$
\es{spectralMu}{
\mu_s(u)
= u \tanh(\pi u) { \pi \left[ u^2 + \left(s + {d - 2 \over 2} \right)^2 \right] \over \left[ 2^{d-1} \Gamma \left( { d+ 1 \over 2} \right) \right]^2} \prod_{j = 1/2}^{(d-4)/2} (u^2 + j^2)  \,.
}
Note that $u^2 + \nu^2$
is the eigenvalue of the STT eigenfunction that diagonalizes the Laplacian $-\nabla^2+\kappa^2$. Here $\Delta$ is the dimension of the dual CFT operator, which according to the structure of the kinetic operators in (\ref{Zs}) is $\Delta_s^{ph} = d + s - 2$ for the spin $s$ gauge field and $\Delta_s^{gh} = d + s - 1$ for the corresponding spin $s-1$ ghost.\footnote{These are the values corresponding to standard boundary conditions. The case of alternate boundary conditions for the spin $s$ gauge fields, leading to conformal higher spin gauge theories at the boundary, was studied in \cite{Giombi:2013yva}.}
The regularized volume of Euclidean $\text{AdS}_{d+1}$ is~\cite{Diaz:2007an,Casini:2011kv,Casini:2010kt}
\es{regAdS}{
{\rm vol}(\text{AdS}_{d+1})= \left\{
\begin{array}{cc}
 \pi^{d/2} \Gamma\left( - {d \over 2} \right) \,, & \text{$d$ odd} \\
{2 (-\pi)^{d/2} \over \Gamma\big(1 + \frac{d}{2} \big) } \log R \,, & \text{$d$ even}
\end{array} \right.
}
 and the volume of the unit $d$-sphere is ${\rm vol}(S^d)=\frac{2\pi^{(d+1)/2}}{\Gamma[(d+1)/2]}$. Finally, the spin factor
 \es{spinFactor}{
 g(s) = {(2s+d-2)(s+d-3)! \over (d-2)!s!} \,, \qquad d \geq 3 \,,
 }
is the number of degrees of freedom of a STT spin s field.

In terms of the spectral zeta-function, the contribution to the one-loop free energy $F^{(1)}=-\log Z^{(1)}$ of a bulk spin $s$ STT field in $\text{AdS}_{d+1}$, with dual dimension $\Delta$, is
\es{FDelta1}{
F_{(\Delta,s)}^{(1)} = - {1 \over 2} \zeta'_{(\Delta,s)}(0) - \zeta_{(\Delta,s)}(0) \log( \ell\, \Lambda) \,,
}
where $\ell$ is the AdS radius of curvature, which we typically set to 1, and $\Lambda$ is the UV cut-off.  The coefficient $\zeta_{(\Delta,s)}(0)$ of the logarithmic term vanishes for even boundary dimension $d$ but is non-zero for odd $d$. 
For the Vasiliev theory in $\text{AdS}_{d+1}$, the full expression for the one-loop correction to the free energy may be written as (\ref{HSZ}), where the higher-spin spectral zeta function is
\es{HSDeffnew}{
\zeta_\text{HS}(z) = \zeta_{(\Delta_0,0)}(z) + \sum_{s = 1}^\infty \big( \zeta_{(\Delta_s^{ph},s)}(z) -  \zeta_{(\Delta_s^{gh},s-1)}(z) \big) \,.
}
Taking $\Delta_0 = d - 2$ boundary condition for $s=0$ corresponds to the $U(N)$ or $O(N)$ singlet sector of the free $d$-dimensional
scalar CFT. For $d>3$, the $s=0$ term does not need to be separated out explicitly, and we have
\es{HSDeffnew}{
\zeta_\text{HS}(z) = \sum_{s =0 }^\infty \big( \zeta_{(s+d-2,s)}(z) -  \zeta_{(s+d-1,s-1)}(z) \big) \, .
}
The subtraction does not affect the contribution of $s=0$, since $\zeta_{(d-1,-1)}(z)$ formally vanishes because $g(-1)=0$ for $d>3$.

Near $z = 0$, the higher-spin spectral zeta function has new power law divergences that are absent in the theory of each individual spin, coming from the infinite sum over all spins. As explained in the introduction, we will show that a natural way to regularize these divergences is through analytic continuation in $z$.
Even when considering a field of given spin in the bulk, it is necessary to evaluate $\zeta_{(\Delta, s)}(z)$ near $z = 0$ by analytic continuation in $z$.  That is, one first integrates~\eqref{spectralZeta} at large $z$ where the integral is convergent and then evaluates the resulting expression near $z = 0$.  Since $u^2 + \nu^2$  is the eigenvalue of the spin $s$ STT Laplacian, this regularization method protects the symmetries of the Laplacian.  This
is analogous to the zeta function regularization method used to, say, evaluate partition functions on the sphere (see, for example,~\cite{Klebanov:2011gs}).
We propose that it is also natural to use $z$ to regulate the sum over spins in the higher-spin theories with infinite towers of higher-spin modes.  That is, we first evaluate~\eqref{HSDeff} at large values of $z$ so that the spin sum and $u$ integrals are convergent.  Then, we analytically continue the resulting, finite expression to evaluate $\zeta_\text{HS}(z)$ near $z = 0$.

\section{Calculations in even $d$}

When the boundary theory is even dimensional, the $d$-sphere free energy is dominated by the conformal anomaly term proportional to $a \log R$, where $a$ is the $a$-type Weyl anomaly coefficient.  Thus, by calculating the $\log R$ term in $F^{(1)}$ in AdS$_{d+1}$, we may calculate the one-loop correction to the $a$ anomaly coefficient, which we call $a^{(1)}$. For even $d$,
$\zeta_{(\Delta,s)}(0)$ vanishes for each spin individually. 
Thus, the contribution to $a^{(1)}$ comes from $\zeta'_\text{HS}(0)$.  The $\log R$ dependence in this expression arises simply from the regularized volume of AdS$_{d+1}$~\eqref{regAdS}.

This calculation has interesting applications to conformal higher spin (CHS) theories that have local higher-derivative actions in
even dimensions $d$ \cite{Fradkin:1985am}. Interacting theories coupling infinite sets of such fields have been proposed \cite{Tseytlin:2002gz,Segal:2002gd}.
Using the relation of CHS fields in $d$-dimensions to massless higher spin fields in AdS$_{d+1}$ with alternate boundary conditions, it is possible to show \cite{Giombi:2013yva,Giombi:2013fka}
that the $a$-coefficient of the CHS theory is $a^{\rm CHS}= - 2 a^{(1)}$. This is because $\zeta'_{(\Delta,s)}(0)$ is odd under $\Delta\rightarrow d-\Delta$
in even $d$.

By calculating the higher spin zeta function explicitly in even $d$, we will be able to show that for the theory containing each integer spin once, $a^{(1)}=0$.
This also implies that for such CHS theories the total $a$-coefficient vanishes: $a^{\rm CHS}=0$. In particular, this resolves the issue about the consistency of the $d=6$ CHS theory raised in \cite{Tseytlin:2013fca}. For the minimal theory containing even spins only, we find in all even $d$ that the regularized correction to the
$a$ anomaly equals that of a real conformal scalar: $a_{\rm min}^{(1)}= a_{S}$. This implies that the total anomaly of the even spin CHS theory coupled to
two real conformal scalars vanishes:
\begin{equation}
a_{\rm min}^{\rm CHS}+ 2 a_{S} = 0\ .
\end{equation}

\subsection{AdS$_5$}

As a first illustration, let us consider the $d = 4$ case, i.e. higher-spin theory in $AdS_5$.  To evaluate $\zeta_\text{HS}(z)$, we must evaluate
\es{zetaHS4}{
{\zeta_\text{HS}(z) \over \log R} = &\lim_{\nu \to 0} {1 \over 12 \pi} \int_0^\infty d u \, u^2 { 1 + u^2 \over (u^2 + \nu^2 )^z } + {1 \over 12 \pi} \int_0^\infty d u \, u^2 \sum_{s=1}^\infty (s+1)^2 { u^2 + (s+1)^2 \over (u^2 + s^2)^z} \\
&- {1 \over 12 \pi} \int_0^\infty d u \, u^2 \sum_{s=1}^\infty s^2 { u^2 + s^2 \over \big(u^2 + (s+1)^2\big)^z} \,.
}
The first term above is subtle.  The result in~\eqref{spectralZeta} is strictly only correct when $\nu > 0$.  One way to obtain the result with $\nu \leq 0$ is to first perform the integral assuming $\nu > 0$ and then analytically continue in $\nu$~\cite{Giombi:2013yva}.  This method shows that the fields with
 $\nu=0$ do not contribute to $\zeta_\text{HS}(z)$ near $z = 0$. In particular, the scalar field in AdS$_5$ does not contribute because it has $\Delta=2$ and
 therefore $\nu=0$.

The spin sums in~\eqref{zetaHS4} may be evaluated explicitly for large enough $z$. For example,
\es{spinSumEx}{
& \int_0^\infty d u \, u^2 \sum_{s=1}^\infty (s+1)^2 { u^2 + (s+1)^2 \over (u^2 + s^2)^z} \\
 &=  \int_0^\infty d u \, u^2 \sum_{s=1}^\infty s^{3-2\, z} (s+1)^2 {s^2\, u^2 + (s+1)^2 \over (u^2 + 1)^z}  \\
 &= \big[ \zeta(2z - 7) + 2 \zeta(2z - 6) + \zeta(2z - 5) \big] \int_0^\infty du \, {u^4 \over (1 + u^2)^z} \\
 &+ \big[ \zeta(2z - 7) + 4 \zeta(2z - 6) + 6 \zeta(2z - 5) + 4 \zeta(2z - 4) + \zeta(2z - 3) \big] \int_0^\infty du \, {u^2 \over (1 + u^2)^z} \,.
 }
 Note that in going between the first and second line above we changed variables $u \rightarrow s^{-1} u$.  Then, using the formula
 \es{pInt}{
 \int_0^\infty d u \, {u^{2p} \over (1 + u^2)^z} = {\Gamma\left( p + \frac12 \right) \over 2} {\Gamma\left(z - p - \frac12 \right) \over \Gamma(z)}
 }
 and performing an analogous computation to~\eqref{spinSumEx} for the ghost contribution to $\zeta_\text{HS}(z)$, we find
 \es{zetaHS4: 1}{
 \frac{\zeta_\text{HS}(z)}{\log R} = {\Gamma\left( z - \frac52 \right) \over 24 \sqrt{\pi} \Gamma(z)} \big[ (4 z - 7) \zeta(2z - 6) + 2 (2z - 5) \zeta(2z - 4)  \big] \,.
 }
 Expanding this expression near $z = 0$, we see that $\zeta_\text{HS}(z) \sim O(z^2)$.  This implies that $a^{(1)} = 0$ in this theory.

 It is also interesting to consider the minimal theory that contains even-spin fields only. The expression for $\zeta_\text{min~HS}(z)$ is obtained from (\ref{zetaHS4}) by restricting the sum to run over the even spins. A straightforward computation yields the final result 
\es{}{
\frac{\zeta_\text{min~HS}(z)}{\log R}&=\frac{2^{-2 z}\Gamma\left(z-\frac{5}{2}\right)}{96\sqrt{\pi} \Gamma(z)}
\big[2^{2 z+1} (4 z-7) \zeta(2z-6)+4^{z+1} (2 z-5) \zeta(2z-4) \\
&-2 \left(4^z-256\right) (z-1) \zeta(2z-7)-3\left(4^z-64\right) (4 z-9) \zeta(2z-5) \\
&-\left(4^z-16\right)(2 z-5) \zeta(2z-3)\big].
}
Expanding this result around $z = 0$, we find that in this minimal theory with even spins only
\begin{equation}
\zeta_\text{min~HS}(z)  = - {z \over 45} \log R + O(z^2).
\end{equation}
This implies that $a^{(1)} = 1 / 90$, which is exactly the $a$ anomaly coefficient of a real scalar field.

\subsection{AdS$_7$}

As a second example, let us consider the higher spin theories in $AdS_7$.  In this case we have
 \es{zetaHS6}{
{\zeta_\text{HS}(z) \over \log R} = &- {1 \over 360 \pi} \int_0^\infty d u \, u^2 {  u^2 + 4 \over (u^2 + 1 )^{z-1} } \\
&- {1 \over 4320 \pi} \int_0^\infty d u \, u^2 (u^2 + 1) \sum_{s=1}^\infty (s+1)(s+2)^2(s+3) { u^2 + (s+2)^2 \over \big(u^2 + (s+1)^2\big)^z} \\
&+ {1 \over 4320 \pi} \int_0^\infty d u \, u^2 (u^2 + 1) \sum_{s=1}^\infty s(s+1)^2(s+2) { u^2 + (s+1)^2 \over \big(u^2 + (s+2)^2\big)^z} \,.
}
Note that unlike in the $d=4$ case, in $d = 6$ it is not necessary to analytically continue in $\nu$ for the scalar mode contribution, since $\nu=1$ ($\Delta=4$).  It is straightforward to evaluate the expression above using the same tools given in the $d = 4$ case.  Including all integer spin fields gives the result
 \es{zetaHS6: 1}{
 \frac{\zeta_\text{HS}(z)}{\log R} &=- {\Gamma\left( z - \frac72 \right) \over 17280 \sqrt{\pi} \Gamma(z)} \big[ 3 (6 z - 11) \zeta(2z - 10) + 6 (2z^2 - 2 z  - 15) \zeta(2z - 8)  \\
 &+2 (2z - 7) (8z - 17) \zeta(2z - 6) + (2z - 7) (2z - 5) \zeta(2z - 4)  \big] \,.
 }
 As expected, near $z = 0$ we find $\zeta_\text{HS}(z) \sim O(z^2)$. Thus, our higher spin zeta function regularization confirms in $d=6$ that $a^{(1)}=0$, in agreement with the expected duality to free scalar theory. At the same time, this also resolves a puzzle about the $a$-anomaly of the $d=6$
 CHS theory raised in \cite{Tseytlin:2013fca}. For the minimal theory with only even spin fields, a direct calculation yields the result
\begin{equation}
\zeta_\text{min~HS}(z) = {z \over 378} \log R + O(z^2),
\end{equation}
which implies $a_{\rm min}^{(1)} = - 1/ 756$.  This is exactly the $a$ anomaly coefficient of a real conformal scalar field in $d = 6$, normalized so that
$F= a \log R$.

\subsection{AdS$_9$}

Let us consider the higher spin theories in $AdS_9$ as an additional explicit example. In this case, the full higher spin zeta function is given by
\es{HS8}{
&\!\!\!{\zeta_\text{HS}(z) \over \log R} =  \int_0^\infty {d u \, u^2 \over 20160 \pi} {  (u^2 + 1)(u^2+9) \over (u^2 + 4)^{z-1} } \\
&\!\!\!+  \int_0^\infty {d u \, u^2 (u^2 + 1)(u^2+4) \over 7257200 \pi} \sum_{s=1}^\infty (s+1)(s+2)(s+3)^2(s+4)(s+5) { u^2 + (s+3)^2 \over \big(u^2 + (s+2)^2\big)^z} \\
&\!\!\!- \int_0^\infty {d u \, u^2 (u^2 + 1)(u^2+4) \over  7257200 \pi} \sum_{s=1}^\infty s(s+1)(s+2)^2(s+3)(s+4) { u^2 + (s+2)^2 \over \big(u^2 + (s+3)^2\big)^z}  \,.
}
Evaluating the integral and the sum over all integer spins, we obtain the result
\es{}{
 &\frac{\zeta_\text{HS}(z)}{\log R} ={\Gamma\left( z - \frac92 \right) \over 58060800 \sqrt{\pi} \Gamma(z)} \big[ 15(8 z - 15) \zeta(2z - 14) + 120z (2z-9) \zeta(2z - 12)  \\
 &+(128z^3-420z^2-3236z+11088)\zeta(2z - 10) \\
 &+4 (2 z-9) (52z^2-417z+755) \zeta(2z -8) -(2 z-9) (2 z-7) (112 z-199)\zeta(2z-6) \\
 &-12 (2 z-9) (2 z-7) (2 z-5) \zeta(2z-4)\big]\,.
}
Expanding around $z=0$, we find $\zeta_\text{HS}(z) \sim O(z^2)$, which implies $a^{(1)}=0$ as expected, in agreement with the duality to the $d=8$ free scalar theory. By restricting the sums in (\ref{HS8}) to even spins only, we can analogously compute the full zeta function in the minimal higher spin theory. Expanding the final result near $z=0$, we find in this case
\begin{equation}
\zeta_\text{min~HS}(z) = -{23z \over 56700} \log R + O(z^2)\,.
\end{equation}
This implies that $a_{\rm min}^{(1)} = 23/ 113400$, which is indeed equal to the $a$ anomaly coefficient of a real conformal scalar in $d=8$.

It is straightforward to generalize these computations to higher even dimensions $d$ and the same conclusions continue to hold.  In the theory with all integer spins $a^{(1)} = 0$, while in the theory with only even spins $a_{\rm min}^{(1)} = a_S$ is the $a$ anomaly coefficient of a single real scalar field.   For the minimal theory, we then learn that $G_N^{-1} a^{(0)} = N a_S - a_S$, where $a^{(0)}$ is the leading contribution to the $a$ anomaly coefficient. Thus, we find that consistency with the duality to the $O(N)$ singlet sector of the free scalar theory always appears to require the shift $G_{N}^{-1}\sim N-1$.

\subsection{AdS$_3$}

 We now perform a similar central charge matching for higher spin theory in AdS$_3$. Here we do not consider the case of the Gaberdiel-Gopakumar duality \cite{Gaberdiel:2010pz,Gaberdiel:2012uj}, but rather the higher spin theory whose spectrum is dual to the theory of $N$ free scalar fields in $d=2$, similarly to the cases we discuss in other dimensions.\footnote{For a one-loop test of the Gaberdiel-Gopakumar duality, see \cite{Giombi:2013fka}.}  In this case, we must be careful when evaluating the contribution from the scalar and vector fields to $\zeta_\text{HS}(z)$.  As was pointed out around~\eqref{zetaHS4}, the result in~\eqref{spectralZeta} is strictly only valid when $\nu > 0$.  We use analytic continuation in $\nu$ to compute contributions with $\nu \leq 0$.  In $d = 2$, we have $\mu_s(u) = u^2 + s^2$, and $g(0) = 1$, $g(s) = 2$ for $ s \geq 1$.  The scalar modes then require the integral
 \es{scalard2}{
 {\zeta_{(\nu + 1, 0)}(z) \over \log R}=- {1 \over \pi} \int_0^\infty d u { u^2 \over ( u^2 + \nu^2)^z } = - { \nu^{3-2 z} \Gamma\left( z - \frac32 \right) \over 4 \sqrt{\pi} \Gamma(z) } \,.
 }
The physical scalar mode has $\Delta=0$ ($\nu=-1$), and its contribution should be defined by analytic continuation from positive $\nu$.
The ``physical" $s=1$ field has $\Delta = 1$ ($\nu = 0$) and does not contribute to $\zeta_\text{HS}(z)$, while its scalar ghost with
$\Delta = 2$ ($\nu = 1$) does contribute.

To calculate the contribution from the modes with $s \geq 2$, we may use the same method as in the higher-$d$ cases, and we find
\es{zetaHS2}{
{\zeta_\text{HS}(z) \over \log R} &={ \big( 1 + (-1)^{-2 z} \big) \Gamma\left( z - \frac32 \right) \over 4 \sqrt{\pi} \Gamma(z ) } - {2 \over \pi} \int_0^\infty d u \sum_{s = 2}^\infty \left[ { u^2 + s^2 \over ( u^2 + (s - 1)^2)^z }- { u^2 + (s-1)^2 \over ( u^2 + s^2)^z } \right] \\
&= - {4 \Gamma\left( z - \frac12 \right) \zeta(2z - 2) \over \sqrt{\pi} \Gamma(z)} - {\Gamma\left(z - \frac32 \right) \over 4 \sqrt{\pi} \Gamma(z)} \big( 1 - (-1)^{-2z} \big)  \,.
}
Near $z = 0$, we find the expected result: $\zeta_\text{HS}(z) \sim O(z^2)$.  If we consider the truncation to the theory with only even spin fields, then instead we find $\zeta_\text{min~HS}(z) \sim {2 z \over 3} \log R + O(z^2)$.  Then, using the normalization for the $c$ anomaly where $F = - {c \over 3} \log R$, we see that the first correction to the $c$ anomaly, $c_{\rm min}^{(1)} = 1$, is exactly that of a real scalar field, in agreement with the results found in other dimensions.

The contribution to $c$ from the bulk field of spin $s$ and its associated spin $s-1$ ghost is
\es{Cs}{
c^{(1)}_0 = 1/2\,, \qquad c^{(1)}_1 = 1/2 \,, \qquad c^{(1)}_s = \big[ 1 + 6 \, s\, (s - 1) \big]  \quad (s \geq 2) \,.
}
In each case, we find the relation $-2c^{(1)}_s=c^{\rm CHS}_s$, where $c^{\rm CHS}_s$ is the central charge of the $d=2$ conformal spin $s$ theory
calculated in \cite{Giombi:2013yva} using AdS$_3$ methods.

\subsection{Alternate regulators} \label{sec: alt}

In general, the integrals that enter into the calculation of $\zeta_{(\Delta,s)}(z)$~\eqref{spectralZeta} in odd dimensional AdS spaces are of the form
\es{Iexample}{
I_\text{ex}(\nu) = \int_0^\infty d u \, { u^{p} \over \big[ u^2 + \nu^2 \big]^z } \,, \qquad p \in \mathbb{Z}_{\geq 0} \,.
}
By rescaling $u$, we may completely remove the $\nu$ dependence from~\eqref{Iexample} and write $I_\text{ex}(\nu) = \nu^{p+1 - 2 z} I_\text{ex}(1)$.  At a practical level, this means that an equivalent way of regulating $\zeta_\text{HS}(0)$ (or also $\zeta'_\text{HS}(0)$) is
\es{altReg: 1}{
\zeta_\text{HS}(0) &= \zeta_{(d - 2, 0)}(0) + \lim_{\alpha \to 0} \sum_{s = 1}^\infty \left( s + {d \over 2}  - 2 \right)^{- \alpha} \zeta_{(d + s - 2, s)}(0) \\
&- \lim_{\alpha \to 0} \sum_{s = 1}^\infty \left( s+  {d \over 2}  - 1 \right)^{- \alpha} \zeta_{(d + s - 1, s-1)}(0) \,.
}
That is, first we use the spectral zeta function to calculate $\zeta_{(\Delta, s)}(0)$ or $\zeta_{(\Delta, s)}'(0)$ for each higher spin field.  Then, we sum the resulting expressions using a zeta function regulator, where the physical modes are summed with $ \left( s + {d \over 2}  - 2 \right)^{-\alpha}$ and the ghost modes with $ \left( s + {d \over 2} - 1 \right)^{-\alpha}$.  After performing the sums, we take the limit $\alpha \to 0$.

It is an interesting and non-trivial observation that~\eqref{altReg: 1} is equivalent to using the regulator scheme
\es{altReg: 2}{
\zeta_\text{HS}(0) = &\zeta_{(d - 2, 0)}(0) + \lim_{\alpha \to 0} \sum_{s = 1}^\infty \left(s +  {d - 3 \over 2} \right)^{- \alpha} \left( \zeta_{(d + s - 2, s)}(0) - \zeta_{(d + s - 1, s-1)}(0) \right) \,.  \\
}
This scheme may also be used to calculated $\zeta_\text{HS}'(0)$, with the obvious substitutions.  We show the equivalence between these two regulators in a variety of examples.  The regulator~\eqref{altReg: 2} may be interpreted, in some sense, as that obtained by taking the ``average" of the $\nu$ coming from the physical modes and the $\nu$ coming from the ghosts.  The regulator~\eqref{altReg: 2} is often the easiest to use in practice.  However, it may be verified in examples that the different ways of regulating the spin sums give equivalent results.\footnote{Another easy to use regulator is to insert
$\exp \big (-\epsilon \left (s+ \frac {d-3} {2} \right ) \big)$ into the sum over $s$, expand the result for small
$\epsilon$, and keep the term of order $\epsilon^0$. In all examples, we find that it is equivalent to the regulator (\ref{altReg: 2}).}

To illustrate the utility of the different regulators, let us revisit the higher spin theories in AdS$_5$.  A straightforward calculation leads to the result
\es{F1s4}{
\zeta'_{( \nu + 2,s)}(0) = {(s+1)^2 \nu^3 \over 180} \big( 5(s+1)^2 - 3 \nu^2 \big) \log R \,.
}
We may then calculate $a^{(1)}$ in the theory with all integer spins using the regulator~\eqref{altReg: 2}, except we replace $(d-3)/2$ by $x$:
\es{zeta4Alt: 1}{
a^{(1)} &= -\lim_{\alpha \to 0} \sum_{s = 1}^\infty (s + x)^{-\alpha} {s^2 \over 360} (1+s)^2 \left[ 3 + 14 \, s(s+1) \right]
={x^3 \left( x - \frac12 \right) (x-1)^3 \over 180} \,.
}
As expected, taking $x= (d-3)/2=\frac12$ causes $a^{(1)}$ to vanish.\footnote{Using this regulator with $x=1/2$ also renders the sum over $c$-anomalies in the
$d=4$ CHS theory \cite{Tseytlin:2002gz,Segal:2002gd} vanishing for any value of the parameter $r^{(b)}$ introduced in
eq. (5.5) of \cite{Tseytlin:2013jya}. We also find that for $r^{(b)}= -1$, which gives
$c_s  = \frac {1} {90} s(1+s) (-4-17s -3s^2+28 s^3+14 s^4)$,
this regulator makes the sum of $c$-anomalies over even spins equal $-1/15$, which is minus that of
a complex scalar. These results are then analogous to our results for the $a$-anomalies in the CHS theories.}   In the minimal theory with only even spins, we calculate
 \es{as41T}{
a^{(1)}_\text{min} =  {1 \over 90} +  {x^3 \left( x - \frac12 \right) (x-1)^3 \over 360} \,,
}
which equals the $a$ anomaly coefficient of a real scalar field when $x = 1/2$ (and also $x=0,1$).

In $d = 6$ (AdS$_7$) we calculate
\es{F1s6}{
\zeta'_{(\nu +3 ,s)}(0) = &{(s+1) (s+2)^2(s+3) \nu^3 \over 453600} \big[ - 35 (s+2)^2+21 \big(5 + s(s+4) \big) \nu^2 - 15 \nu^4 \big] \log R \,.
}
The contribution to the $a$-anomaly from the spin $s$ bulk field and its associated spin $s-1$ ghost is
\es{as6}{
 a^{(1)}_s =  { \left( \gamma_6 - \frac14 \right)^2 \over 9676800} \big( 25 - 884 \gamma_6 + 2288 \gamma_6^2 - 704 \gamma_6^3 \big)   \,, \qquad \gamma_6 = \left( s + \frac32 \right)^2 \,.
}
As usual, this is $-1/2$ of the Weyl anomaly of the conformal spin $s$ theory in $d=6$ that was calculated in \cite{Giombi:2013yva,Tseytlin:2013fca}.\footnote{
We note that (\ref{as6}) applies for all integer spins including $s=0$; an analogous result holds in all $d\geq 4$ since the formal subtraction of the spin $-1$ ghosts does not affect the result. In $d=6$, $a^{(1)}_0={1 \over 1512}$ which implies that the conformal spin $0$ theory has
$a^{\rm CHS}_0=-2 a^{(1)}_0 = -{1 \over 756}$. This is the
$a$-anomaly of a standard conformally coupled scalar; indeed, in $d=6$ the CHS action for $s=0$ has two derivatives and is the standard conformal scalar action.
In general, the kinetic term for a spin $s$ CHS field has $d-4+ 2s$ derivatives \cite{Fradkin:1985am}.}
Using a modified form of the regulator~\eqref{altReg: 2}, where $(d-3)/2$ is replaced by $x$, leads to the result
\es{a16}{
a^{(1)} &=  \lim_{\alpha \to 0} \sum_{s=0}^\infty a^{(1)}_s (s+x)^{-\alpha}\\
 &= { (x-1)^3 \left( x - \frac32 \right) (x-2)^3 \over 453600} \big( 3 x^4 - 18 x^3 + 29 x^2 - 6x - 6 \big)  \,.
}
In the minimal even spin theory we instead find
\es{a16T}{
a^{(1)}_\text{min} &=  \lim_{\alpha \to 0} \sum_{s=0, 2, \dots}^\infty a^{(1)}_s (s+x)^{-\alpha}\\
 &= - {1 \over 756} + { (x-1)^3 \left( x - \frac32 \right) (x-2)^3 \over 907200} \big( 3 x^4 - 18 x^3 + 29 x^2 - 6x - 6 \big)  \,.
}
Remarkably, we find that when we choose $x = 1$, $3/2$, $2$ such that $a^{(1)} = 0$, $a^{(1)}_\text{min}$ is exactly the $a$ anomaly coefficient of a real scalar field.
In general even dimension $d$, we have verified explicitly (through $d = 20$) that choosing $x = (d-4)/2, (d-3)/2, (d-2)/2$ leads to $a^{(1)} = 0$ and $a^{(1)}_\text{min} = a_S$, where $a_S$ is the $a$ anomaly coefficient of a real scalar field.

\section{Calculations in odd $d$}

The even dimensional AdS theories differ from the odd dimensional ones in two key ways.  First, $\zeta_\text{HS}(0)$ does not vanish trivially in this case.  Indeed, in even dimensional space-time there is a logarithmic divergence in the one-loop free energy (for conformal fields, this is related to the conformal anomaly), and each higher spin field contributes to this quantity. Thus, the logarithmic term involving $\zeta_\text{HS}(0)$ may vanish only after performing the properly regularized sum over all spins.  Second, the spin $s$ spectral densities $\mu_s(u)$ are not pure polynomials in $u$.  This is because of the factor $\tanh(\pi u)$ in~\eqref{spectralMu}.  We must work harder in the odd $d$ theories to calculate $\zeta_\text{HS}(0)$ and $\zeta_\text{HS}'(0)$. In the following subsections we work out some explicit examples.

\subsection{AdS$_4$}

The simplest odd-dimensional example is $d = 3$, where we may use the identity $\tanh(\pi u) = 1 - 2 (1 + e^{2 \pi u})^{-1}$ to write
\es{zetaHS0d3: 1}{
&\zeta_\text{HS}(z)  =  {1 \over 6}  \lim_{\nu \to - 1/2} \left[ \int_0^\infty d u \, u { \left( u^2 + \frac14 \right) \over (u^2 + \nu^2)^z} - 2  \int_0^\infty {d u \, u \over (1 + e^{2 \pi u} )} { \left( u^2 + \frac14 \right) \over (u^2 + \nu^2)^z} \right]  \\
&+\sum_{s = 1}^\infty \int_0^\infty {d u \, u \over 3} \left[  { \left( s + \frac12 \right)}   {\left[ u^2 + \left(s + \frac12 \right)^2 \right]  \over \left[ u^2 + \left(s- \frac12 \right)^2 \right] ^z} -    { \left( s - \frac12 \right) }  {\left[ u^2 + \left(s - \frac12 \right)^2 \right]  \over \left[ u^2 + \left(s+ \frac12 \right)^2 \right] ^z} \right]  \\
&- 2\sum_{s = 1}^\infty \int_0^\infty {d u  \,u \over 3 (1 + e^{2 \pi u} )} \left[  { \left( s + \frac12 \right)}   {\left[ u^2 + \left(s + \frac12 \right)^2 \right]  \over \left[ u^2 + \left(s- \frac12 \right)^2 \right] ^z} -    { \left( s - \frac12 \right) }  {\left[ u^2 + \left(s - \frac12 \right)^2 \right]  \over \left[ u^2 + \left(s+ \frac12 \right)^2 \right] ^z} \right]  \,.
}
The second line above involves pure powers of $u$, and it may be evaluated using the methods presented in the even $d$ section.  Similarly, it is straightforward to evaluate the first term in the scalar contribution in the first line.  These terms contribute (near $z = 0$)
\es{zetaHS0d3: 2}{
&  {1 \over 6}  \lim_{\nu \to - 1/2} \int_0^\infty d u \, u { \left( u^2 + \frac14 \right) \over (u^2 + \nu^2)^z} \\
&+\sum_{s = 1}^\infty \int_0^\infty {d u \, u \over 3} \left[  { \left( s + \frac12 \right)}   {\left[ u^2 + \left(s + \frac12 \right)^2 \right]  \over \left[ u^2 + \left(s- \frac12 \right)^2 \right] ^z} -    { \left( s - \frac12 \right) }  {\left[ u^2 + \left(s - \frac12 \right)^2 \right]  \over \left[ u^2 + \left(s+ \frac12 \right)^2 \right] ^z} \right]  \\
&=
\left( - {\zeta(3) \over 8 \pi^2} + {75 \zeta(5) \over 64 \pi^4} \right) z + O(z^2) \,.
}

Evaluating the terms with $(1 + e^{2 \pi u})$ in the denominator requires more work.  First, we focus on $\zeta_\text{HS}(0)$.  Setting $z = 0$, we may perform the integral in the second term of the first line of~\eqref{zetaHS0d3: 1}:
\es{firstTerm: 1}{
\int_0^\infty {d u \, u \over (1 + e^{2 \pi u} )} { \left( u^2 + \frac14 \right)} = {17 \over 1920}  \,.
}
 To evaluate the third line in~\eqref{zetaHS0d3: 1}, we  first set $z = 0$ and perform the integral over $u$.  Then, we may evaluate the spin sum using the Hurwitz zeta-function regularization (\ref{altReg: 1}). We insert $\left( s - {1 \over 2} \right)^{-\alpha}$ for the physical mode contribution and $\left( s + {1 \over 2} \right)^{-\alpha}$ for the ghosts.  After performing the spin sum, we take the limit $\alpha \to 0$:
 \es{thirdTerm: 1}{
&\lim_{\alpha \to 0} \sum_{s = 1}^\infty \int_0^\infty {d u  \,u \over 3 (1 + e^{2 \pi u} )}   { \left( s + \frac12 \right)^{1 - \alpha}} \  {\left[ u^2 + \left(s + \frac12 \right)^2 \right]  }  \\
&-  \lim_{\alpha \to 0} \sum_{s = 1}^\infty \int_0^\infty {d u  \,u \over 3 (1 + e^{2 \pi u} )}  { \left( s - \frac12 \right)^{1 - \alpha} }  {\left[ u^2 + \left(s - \frac12 \right)^2 \right] = -{ 17 \over 11520} \,. }
 }
Combined together, the results in~\eqref{zetaHS0d3: 2},~\eqref{firstTerm: 1}, and~\eqref{thirdTerm: 1} show that $\zeta_\text{HS}(0) = 0$. 

Evaluating $\zeta_\text{HS}'(0)$ is technically more involved than the calculation of $\zeta_\text{HS}(0)$.  We will show multiple different methods to perform this calculation, all of which lead to the same results.  To begin, we may calculate the contribution of the second term (originating from the scalar mode) in the first line of~\eqref{zetaHS0d3: 1}.  With a bit of effort, the relevant integral may be performed analytically:
  \es{integ: 1}{
  \int_0^\infty {d u \, u \over (1 + e^{2 \pi u} )} { \left( u^2 + \frac14 \right)} \log \left(u^2 + \frac14 \right)  &= {799 \over 11520} - {7 \gamma \over 960} + { \log 2 \over 120}  -{\log A \over 2} - {7 \log \pi \over 960} \\
  &+ {3 \zeta(3) \over 8 \pi^2} + {15 \zeta'(-3) \over 8} + {21 \zeta'(4) \over 32 \pi^4}  \,.
 }

The first method we illustrate for dealing with the spin sums in the third line of~\eqref{zetaHS0d3: 1} is straightforward, though we must numerically calculate finite integrals at the end of the calculation.
First, we take the derivative of that line with respect to $z$ and set $z = 0$.  Then, we regularize the spin sums by writing
\es{Areg}{
\log\left[ u^2 + \left(\Delta - \frac32 \right)^2 \right]  = \lim_{\alpha \to 0} \partial_\alpha \left[ \left( u + i \left(\Delta - \frac32 \right) \right)^\alpha + \left( u - i \left(\Delta - \frac32 \right) \right)^\alpha \right]
}
and performing the sums over $s$ at finite $\alpha$.  Afterwards, we evaluate the $\alpha$ derivative and take the limit $\alpha \to 0$.  For example, consider the term
\es{exampleTerm}{
&\sum_{s= 1}^\infty \int_0^\infty {d u \, u \over (1 + e^{2 \pi u} )} \log\left[ u^2 + \left(s - \frac12\right)^2 \right] \\
&\to \int_0^\infty {d u \, u \over (1 + e^{2 \pi u} )} \lim_{\alpha \to 0} \partial_\alpha \sum_{s=1}^\infty \left[ \left( u + i \left( s - \frac12 \right) \right)^\alpha + \left( u - i \left( s - \frac12 \right) \right)^\alpha \right] \\
&= \int_0^\infty {d u \, u \over (1 + e^{2 \pi u} )} \log\left(1 + e^{-2 \pi u} \right)
={\zeta(3) \over 32 \pi^2} \,.
}
The other terms in the third line of~\eqref{zetaHS0d3: 1} may be evaluated in the same way.  However, unlike in the example above, we are not able to perform all of the final integrals over $u$ analytically.  Performing these integrals numerically, we are able to confirm to over 20 digits of precision that with this regularization scheme
\es{final3Integ}{
\sum_{s = 1}^\infty \int_0^\infty {d u  \,u \over  (1 + e^{2 \pi u} )} &\left[  { \left( s + \frac12 \right)}   {\left( u^2 + \left(s + \frac12 \right)^2 \right)  \log \left( u^2 + \left(s- \frac12 \right)^2 \right)}  \right. \\
&\left.-    { \left( s - \frac12 \right) }  {\left( u^2 + \left(s - \frac12 \right)^2 \right)  \log\left( u^2 + \left(s+ \frac12 \right)^2 \right)} \right] \\
&=- {799 \over 23040} + {7 \gamma \over 1920} - { \log 2 \over 2400}  +{\log A \over 4} + {7 \log \pi \over 1920} \\
& - {225 \zeta(5) \over 128 \pi^4} - {15 \zeta'(-3) \over 16} - {21 \zeta'(4) \over 64 \pi^4}  \,.
}
Combining the results in~\eqref{zetaHS0d3: 2},~\eqref{integ: 1}, and~\eqref{final3Integ}, we then see that $\zeta_\text{HS}'(0) = 0$.

The calculation in the minimal theory proceeds analogously to that presented above, and we find $\zeta_\text{HS}(0) = 0$ and $\zeta_\text{HS}'(0) = - 2 F_S$, where $F_S = {1 \over 2^4} \left( 2 \log 2 - {3 \zeta(3) \over \pi^2} \right)$ is the $S^3$ free energy of a real scalar field.  This implies, as expected, that $F^{(1)}_{\rm min} = F_S$ in this theory, as found in \cite{Giombi:2013fka}.

\subsubsection{Alternate regulators}

Strong consistency checks of the results for $\zeta_\text{HS}(z)$ near $z = 0$ are obtained by evaluating $\zeta_\text{HS}(0)$ and $\zeta_\text{HS}'(0)$ with the alternate regulators presented in Sec.~\ref{sec: alt}.  In particular, we note that $\zeta_\text{HS}(0)$ and $\zeta_\text{HS}'(0)$ were calculated in~\cite{Giombi:2013fka} using the regulator~\eqref{altReg: 2}, and the results presented there agree with those in the previous section.  Below we discuss the regulators~\eqref{altReg: 1} and~\eqref{altReg: 2} in more detail.

We begin with $\zeta_\text{HS}(0)$.  Expressions for $\zeta_{(\Delta,s)}(0)$ may be calculated by using the identity $\tanh(\pi u) = 1 - 2 (1 + e^{2 \pi u})^{-1}$ and analytically continuing in $z$ in~\eqref{spectralZeta}~\cite{Camporesi:1993mz}.
 In $d = 3$, this procedure leads to the result
 \es{d3Zeta0}{
 \zeta_{(\nu + 3/2,s)}(0) = {s + \frac12 \over 12} \left[ \nu^4 - \left( s + \frac12 \right)^2 \left(2 \nu^2 + {1 \over 6} \right) - {7 \over 240} \right] \,.
 }
 Using the regulator~\eqref{altReg: 1}, we then compute
 \es{HS0: alt reg: d3: 1}{
 \zeta_\text{HS}(0) = &-{1 \over 180} + \frac19 \lim_{\alpha \to 0} \sum_{s = 1}^\infty \left( s - \frac12 \right)^{- \alpha} \left( - {3 s^5 \over 4} - {15 s^4 \over 8} + s^3 + {3 s^2 \over 8} - {7 s \over 20} - {1 \over 20} \right) \\
&- \frac19 \lim_{\alpha \to 0} \sum_{s = 1}^\infty \left( s+  \frac12 \right)^{- \alpha}  \left( - {3 s^5 \over 4} + {15 s^4 \over 8} + s^3 - {3 s^2 \over 8} - {7 s \over 20} + {1 \over 20} \right) \\
&=- {1 \over 180} + {769 \over 483840} + {1919 \over 483840}
= 0 \,.
 }
 If instead we regulated the physical mode contribution with $\left( s + x \right)^{-\alpha}$ and the ghost contribution with $\left( s+ y \right)^{-\alpha}$, we would find that $\zeta_\text{HS}(0) = 0$ so long as $x + y = 0$.  This is also true if we consider the minimal theory with only even spin fields. In particular, this result implies as a special case that if we calculate $\zeta_\text{HS}(0)$ using the regulator~\eqref{altReg: 2}, but with $(d-3)/2$ replaced by $x$, we would find that $\zeta_\text{HS}(0)$ if we choose $x=0$. Indeed, the explicit calculation yields
\es{d3SumR1}{
\zeta_\text{HS}(0) &= -{1 \over 180} \left[ 1 + \lim_{\alpha \to 0} \sum_{s = 1}^\infty \left( 2 - 15 s^2 + 75 s^4   \right) (s + x)^\alpha  \right]  \\
&= {x (2 - 5 x^2 + 15 x^4) \over 180}  \,.
}
The only real $x$ that gives vanishing $\zeta_\text{HS}(0)$ is $x = 0$. If we restrict to even spin fields, we find the same result:
\es{d3Even}{
1 + \lim_{\alpha \to 0} \sum_{s = 2, 4, \cdots}^\infty \left( 2 - 15 s^2 + 75 s^4   \right) (s + x)^\alpha = -\frac12 x \big(2 - 5 x^2 + 15 x^4 \big) \,.
}

Now we briefly discuss the calculation of $\zeta_\text{HS}'(0)$.  That calculation was performed in detail in~\cite{Giombi:2013fka} using the regulator~\eqref{altReg: 2}.  We show here that the regulator~\eqref{altReg: 1} gives the same answer.  Both calculations use the result~\cite{Camporesi:1993mz} (see Appendix~\ref{Method})
\es{d3Result}{
\zeta'_{(\nu + 3/2,s)}(0) &= J_{(\nu + 3/2,s)} + K_{(\nu + 3/2,s)} \,,
}
where
\es{JK3}{
J_{(\nu + 3/2,s)} &= {2 \left( s + \frac12 \right) \over 3} \left[ {\nu^2 \over 48} \left( 1 + 6 \nu^2\right) +   \left( s + \frac12\right)^2 c_0 + c_1 \right] \,, \\
 K_{(\nu + 3/2,s)} &=- {2 \left( s + \frac12 \right) \over 3}   \int_0^\nu dx \, x \left[ x^2 - \left( s + {1 \over 2} \right)^2 \right] \psi\left( x + \frac12 \right)\,.
}
Here $\psi(y)=\frac{\Gamma'(y)}{\Gamma(y)}$ is the digamma function, and $c_0$, $c_1$ are $s$-independent constants.  First we concentrate on the contribution from $J_{(\nu + 3/2,s)}$.  Using the regulator~\eqref{altReg: 1}, we calculate
\es{sumd3Ex: 1}{
&J_{(1,0)} +\lim_{\alpha \to 0} \sum_{s = 1}^\infty \left[ \left( s - \frac12 \right)^{-\alpha} J_{(3+s - 2, s)} - \left( s + \frac12 \right)^{-\alpha} J_{(3+s - 1, s - 1)}  \right] \\
&= \left( {5 \over 1152} + {c_0 \over 12} + {c_1 \over 3} \right) + \left( {53 \over 241920} + {113 c_0 \over 1440} + {c_1 \over 36} \right) - \left( {1103 \over 241920} + {233 c_0 \over 1440} + {13 \, c_1 \over 36} \right) \\
&= 0 \,.
}
This sum also vanishes if evaluated using the regulator~\eqref{altReg: 2} and if the sums are taken over even spin fields only.

The contribution from the $K_{(\nu + 3/2,s)}$ is more complicated.  Following~\cite{Giombi:2013fka}, we use the integral representation for $\psi(y)$:
\es{psiy}{
\psi(y) = \int_0^\infty dt \left( {e^{-t} \over t} - {e^{-yt} \over 1 - e^{-t} } \right) \,.
}
Then, we calculate
\es{diffK3}{
&\lim_{\alpha \to 0} \sum_{s=1}^\infty \left[ \left( s - \frac12 \right)^{-\alpha} K_{(s+1 ,s)} -\left( s + \frac12 \right)^{-\alpha} K_{(s+2,s-1)} \right] \\
&= \int_0^\infty dt  \left[ {191 e^{-t} + 1349 e^{-2t} + 1334 e^{-3t} + 202 e^{-4t} - 5 e^{-5t} + e^{-6t} \over 192 (1 - e^{-t} )^5 t }  \right. \\
&\left. + {e^{-{t \over 2}} + 18 e^{-t} + e^{-{3t \over 2}} - 2 e^{-2t} \over 12(1-e^{-t})^2 t^2 } - {3 e^{-t} + 6 e^{-2t} - e^{-3t} \over (1 - e^{-t} )^3 t^3} - 2 {e^{-{t \over 2}} + 3 e^{-t} - e^{-{3t \over 2}} + e^{-2t} \over (1- e^{-t})^2 t^4 } \right] \,.
}
In deriving the expression above, we first performed the integral over $x$ (see~\eqref{JK3}), then we summed over $s$, and then we took the limit $\alpha \to 0$.  This is exactly the expression one finds if instead the sum is performed using the regulator~\eqref{altReg: 2}, as in~\cite{Giombi:2013fka}.

A key point is that~\eqref{diffK3} has only pure power-law divergences in $t$.  That is, the integrand has the expansion ${8 \over 3 t^4} - {1 \over 9 t^2} + O(t^0)$ near $t = 0$.  Had we instead used the regulator~\eqref{altReg: 2} with $(d-3)/2$ replaced by $x$, there would have generically been a $1/t$ term in this expansion.  With the correct choice of regulator, there is not $1/t$ term in the expansion, and we may regulate the pure power-law divergences.  This procedure is described in~\cite{Giombi:2013fka}, where it is shown that~\eqref{diffK3} evaluates to
\es{diffK32}{
\sum_{s=1}^\infty \left[ K_{(s+1 ,s)} -K_{(s+2,s-1)} \right]
= - {11 \over 1152} + {11 \log 2 \over 2880} + {\log A \over 8} - {5 \zeta'(-3) \over 8} - {\zeta'(-2) \over 2} \,.
}
This exactly cancels the contribution from $ K_{(1 ,0)}$, so that in the end $\zeta'_\text{HS}(0) = 0$.

\subsection{AdS$_6$ and beyond} \label{Ads6}

In the previous section, we calculated $\zeta_\text{HS}(0)$ and $\zeta_\text{HS}'(0)$ explicitly in AdS$_4$ using various equivalent regulators.  In this section we primarily focus on $AdS_6$, though we also mention higher dimensions.  For definiteness, we use the regulator~\eqref{altReg: 2} in this section.  However, we have checked that both regulating explicitly in the spectral zeta-function parameter $z$ and using the alternate regulator~\eqref{altReg: 1} give equivalent results.

We begin with the calculation of $\zeta_\text{HS}(0)$ in $AdS_6$.  First, we must calculate $\zeta_{(\Delta,s)}(0)$: 
\es{zeta06}{
\zeta_{(\Delta,s)}(0) &= {1 \over 360} (s+1) \left( s+ \frac32 \right) (s+2) \left[ \lim_{z \to 0} \int_0^\infty du \, u { \left(u^2 + \frac14 \right) \left( u^2 + \left(s + \frac32 \right)^2 \right) \over (u^2 + \nu^2)^z} \right. \\
&\left. - 2 \int_0^\infty du \, u { \left(u^2 + \frac14 \right) \left( u^2 + \left(s + \frac32 \right) \right) \over (1 + e^{2 \pi u})} \right] \,.
}
To evaluate the first integral above, we continue to large enough $z$ so that the integral converges, then after evaluating the integral we take the limit $z \to 0$.  The second integral is convergent, and an explicit evaluation leads to
\es{zeta061}{
\zeta_{(\Delta,s)}(0) &= {1 \over 2160} (s+1) \left( s+ \frac32 \right) (s+2) \left[ \nu^6 - \frac32 \nu^4 \left( s(s+3) + \frac52 \right) \right. \\
&\left. + \frac{3}{4} \nu^2 \left(s + \frac32 \right)^2 + \frac{17}{160} s(s + 3) + \frac{367}{1344} \right]  \,.
}
From this result, we may calculate
\es{d5Sum}{
\zeta_\text{HS}(0) =  {1 \over 1512} \left[ 1 + \sum_{s = 1}^\infty {\gamma_5 \over 20} \left( 6 - 21 \gamma_5 + 98 \gamma_5^2 - 63 \gamma_5^3   \right)  \right]  \,, \qquad \gamma_5 \equiv (s + 1)^2 \,.
}
As in the $d = 3$ case, we may try to regulate~\eqref{d5Sum} with a more general regulator parameterized by $x$:
\es{d5Sum-x}{
 1 +\lim_{\alpha \to 0} &\sum_{s = 1}^\infty {\gamma_5 \over 20} \left( 6 - 21 \gamma_5 + 98 \gamma_5^2 - 63 \gamma_5^3   \right) (s + x)^{-\alpha}  = {(x-1)^3 \over 151200} \big( -24 + 28 x \\
  &+126 x^2 - 420 x^3 + 455 x^4 - 210 x^5 + 35 x^6 \big) \,.
}
The only rational $x$ which causes the expression above to vanish is $x = 1$, which is exactly the value of $x$ that corresponds to the regulator~\eqref{altReg: 2}.  It is also useful to consider the theory with only even spins.  We find the regulated result
\es{d5SumE}{
1+&\lim_{\alpha \to 0} \sum_{s = 2, 4, \cdots}^\infty {\gamma_5 \over 20} \left( 6 - 21 \gamma_5 + 98 \gamma_5^2 - 63 \gamma_5^3   \right) (s + x)^{-\alpha}  =  \\
&{(x-1)^3 \over 302400} \big( -24 + 28 x +126 x^2 - 420 x^3 + 455 x^4 - 210 x^5 + 35 x^6 \big) \,,
}
and this implies that the theory with even spins only is also free of the logarithmic divergence when $x = 1$.

This part of the calculation generalizes easily to higher even dimension AdS spaces.  For example, a direct calculation in AdS$_8$ gives
  \es{zeta081}{
\zeta_{(\Delta,s)}(0) &= {(s+1) (s+2) \left( s+ \frac52 \right) (s+3)(s+4) \over 2419200}  \left[ \nu^8 - \frac43 \nu^6 \left( s(s+5) + \frac{35}{2} \right) \right. \\
&\left. +5\nu^4 \left(s(s+5) + \frac{259}{5} \right) - \frac94 \nu^2 \left( s + \frac52 \right)^2 -  \frac{367}{1008} s(s + 5) -\frac{27859}{11520} \right]  \,,
}
and from this we calculate
\es{d7Sum}{
\zeta_\text{HS}(0) = {127\over 226800} \left[ 1 -  \sum_{s = 1}^\infty {\gamma_7 (\gamma_7 - 1) \over 12192}
\left( 24 - 80 \gamma_7 + 363 \gamma_7^2 - 258 \gamma_7^3 + 39 \gamma_7^4   \right)  \right] \,,
}
where $\gamma_7 = \left( s + 2 \right)^2$.
Evaluating the sum above using the regulator~\eqref{altReg: 2} leads to the result
\es{d7Sum2}{
\zeta_\text{HS}(0)=
{39 \zeta(-12) - 297 \zeta(-10) + 621 \zeta(-8) - 442 \zeta(-6) + 104 \zeta(-4) - 24 \zeta(-2) \over 21772800}
=0 \,.
}
Moreover, it may also be checked that~\eqref{d7Sum} vanishes if the sum is over even spins only.  If we try to instead regulate these sums by inserting $(s+x)^{-\alpha}$ for some $x$ then taking the limit $\alpha \to 0$ after performing the sums, as was done in the $d = 3$ and $d= 5$ cases, we find that the only rational value of $x$ which causes the sums to vanish is $x = 2$, consistent with the regulator~\eqref{altReg: 2}.  We have checked explicitly that these conclusions continue naturally to higher dimensions through AdS$_{30}$, and indeed we expect this procedure to work in all AdS$_d$ with $d$ even.

We now move on to the calculation of $\zeta_\text{HS}'(0)$.  To do this calculation, we need to evaluate $\zeta'_{(\Delta,s)}(0)$.  A method for calculating these functions is described in Appendix~\ref{Method}, and in $AdS_6$ we find the result
\es{d5Resultzp-ads6}{
\zeta'_{(\Delta,s)}(0) = J_{(\Delta,s)} + K_{(\Delta,s)} \,,
}
where
\es{J5}{
J_{(\nu + 5/2,s)} &= -{g(s) \over 60} \left[ \nu^2 { 107 + 40 s (s+3) \over 1920} + \nu^4 {29 + 12 s (s + 3) \over 96} - {\nu^6 \over 8} \right. \\
&\left.  + \frac14 \left( s +\frac32\right)^2 c_0 + \left( s (s + 3) + \frac52 \right) c_1 + c_2 ] \right]
}
and
\es{K5}{
K_{(\nu + 5/2,s)} &= - {g(s) \over 60}  \int_0^\nu dx \, x \left( x^2 - \frac14 \right) \left[ x^2 - \left( s + {3 \over 2} \right)^2 \right] \psi\left( x + \frac12 \right) \,.
}
As in $AdS_4$, it is much easier to calculate the contribution from $J_{(\Delta,s)}$ than from $K_{(\Delta,s)}$.

Let us begin by concentrating on the contribution of $J_{(\Delta,s)}$ to $\zeta_\text{HS}'(0)$.  A first observation is that $g(s) - g(s-1) = \gamma_5$.  Then, using the fact that
\es{gs05}{
g(0) + \sum_{s=1}^\infty \left[  g(s) -  g(s-1)\right] = \zeta(-2) = 0
}
when evaluated with the regulator~\eqref{altReg: 2}, we immediately see that all terms in~\eqref{J5} that do not depend on $s$ beyond the overall $g(s)$ do not contribute to $\zeta_\text{HS}'(0)$.
Moreover, we may use the relation
\es{gs052}{
\frac94 g(0) + \sum_{s = 1}^\infty \left[ \left( s+ \frac32 \right)^2 g(s) - \left( s + \frac12 \right)^2 g(s-1)\right] = \frac94 + \frac{1}{12} \sum_{s=1}^\infty \gamma_5(7 + 20 \gamma_5) = 0  \,,
}
to conclude that the $c_0$ and $c_1$ terms in~\eqref{J5} also do not contribute when the regulator~\eqref{altReg: 2} is used.  Similar calculations may be carried out for the $\nu$-dependent terms in~\eqref{J5}, and the same conclusions are found.

Having shown that the $J_{(\Delta,s)}$ terms do not contribute to $\zeta_\text{HS}'(0)$, let us now concentrate on the $K_{(\Delta,s)}$ contributions.  Following the $AdS_4$ calculation, and using the regulator~\eqref{altReg: 2} (or equivalently~\eqref{altReg: 1}), we find
{\footnotesize
\es{diffK5}{
&\sum_{s=1}^\infty \left( K_{(s+3 ,s)} - K_{(s+4,s-1)} \right) = - \int_0^\infty dt  \left[ { 15 e^{3t} + 5 e^{2t} + 5 e^{t} - 1  \over (e^t-1)^5 t^5} \right. \\
& +{2- 2e^{t/2}-4 e^t+6 e^{3t/2}+10e^{2t}-6e^{5t/2}+2e^{7t/2}\over (e^t-1)^4 t^6}
+ {9 e^{5t} + 241 e^{4t} + 206 e^{3t} + 30 e^{2t} - 7 e^{t} +1 \over 12 (e^t - 1)^7 t^3} \\
& \left. - { e^{t/2} \big(e^{5t} - 5 e^{4t} - 32 e^{7t/2} + 10 e^{3t} - 96 e^{5t/2} - 10e^{2t} - 32 e^{3t/2} + 5 e^t - 1 \big) \over 4 (e^t-1)^6 t^4}  \right. \\
&+ {9 e^{11t/12} - 45 e^{9t/2} - 1360 e^{4t} + 90 e^{7t/2} - 3648 e^{3t} - 90 e^{5t/2} - 1440 e^{2t} + 45 e^{3t/2} + 64 e^{t} -9 e^{t/2} - 16 \over 960 (e^t-1)^6 t^2 }  \\
& \left. -{13 e^{8t} + 22923 e^{7t} + 323028 e^{6t} + 943548 e^{5t} + 946278 e^{4t} + 320922 e^{3t} + 24132 e^{2t} - 468 e^{t} + 117 - 13e^{-t}\over 23040 (e^t-1)^9 t }\right]\,.}}
Near $t = 0$, the integrand has the expansion
\es{diffK5exp}{
 {56 \over 3 t^8} - {11 \over 15 t^6} + {1 \over 60 t^4} - {1 \over 2700 t^2} + (\text{finite at $t = 0$} ) \,.
}
One way of regularizing~\eqref{diffK5} is simply to subtract the power-law divergent terms in~\eqref{diffK5exp} and perform the remaining finite integral.  This may be done numerically, and the result is consistent with
\es{diffK5reg}{
\sum_{s=1}^\infty \left( K_{(s+3 ,s)} - K_{(s+4,s-1)} \right) &= + {1181 \over 1382400} - {211 \log 2 \over 483840} - {23 \log A \over 1920} \\
& + {\zeta(3) \over 96 \pi^2} + {\zeta(5) \over 32 \pi^4} + {7 \zeta'(-3) \over 192} - {21 \zeta'(-5) \over 640} \,.
}

An alternative and more elegant, if slightly more involved, way to derive~\eqref{diffK5reg} exactly is the following.  As in the $d = 3$ calculation~\cite{Giombi:2013fka}, we use the integral representation of the Hurwitz-Lerch function
\es{HurLer}{
\Phi(z,s,v) = { 1 \over \Gamma(s)} \int_0^\infty dt \, {t^{s-1} e^{v t} \over 1 - z e^{-t} } = \sum_{n=0}^\infty (n+v)^{-s} z^n
}
to express~\eqref{diffK5} in terms of $\partial_z^{(p)} \Phi(z,s,v)$, $p \in \mathbb{Z}_{\geq 0}$, evaluated at $z = 0$.
Then, by analytic continuation in $s$, we may relate the resulting sums over $n$ to the Hurwitz zeta function through the relation
\es{Hur}{
\zeta(s,v) = \sum_{n=0}^\infty (n+ v)^{-s}  \,.
}
This way of calculating~\eqref{diffK5} exactly reproduces~\eqref{diffK5reg}.

Now
we calculate the spin 0 contribution:
\es{spin05}{
K_{(3,0)} &=  -{1 \over 60}  \int_0^{1/2} dx \, x \left( x^2 - \frac14 \right) \left( x^2 - \frac94 \right) \psi\left( x + \frac12 \right) \\
&= - {1181 \over 1382400} + {211 \log 2 \over 483840} + {23 \log A \over 1920} - {\zeta(3) \over 96 \pi^2} - {\zeta(5) \over 32 \pi^4} - {7 \zeta'(-3) \over 192} + {21 \zeta'(-5) \over 640} \,.
}
This contribution exactly cancels that from the higher-spin modes in~\eqref{diffK5reg}, showing that $F^{(1)} = 0$ in the AdS$_6$ Vasiliev theory
with all integer spins.

Now we consider the minimal version of Vasiliev's theory in the bulk that only has the even-spin fields.  A straightforward calculation shows that
\es{diffJ5reg2}{
J_{(3 ,0)} + \sum_{s=2,4, \cdots}^\infty \left( J_{(s+3 ,s)} - J_{(s+4,s-1)} \right) = 0
}
when regulated using~\eqref{altReg: 2}.  Then, through a calculation analogous to that above in the theory of all integer spins, we find
\es{diffK52}{
&\sum_{s=2, 4, \cdots}^\infty \left( K_{(s+3 ,s)} - K_{(s+4,s-1)} \right) = \sum_{s=1}^\infty \left( K_{(s+3 ,s)} - K_{(s+4,s-1)} \right) + \delta K_\text{even} \,.
}
Above, the sum over all spins on the right hand side is given by the divergent integral in~\eqref{diffK5}, and we already showed that once regularized this evaluates to $-K_{(3,0)}$.  The remaining contribution, which we call $\delta K_\text{even}$, is in fact a convergent integral over $t$.  Performing this integral, we find
\es{deltaK5even}{
\delta K_\text{even} = -2 F_S  \,, \qquad F_S = - {1 \over 2^8} \left( 2 \log 2 + {2 \zeta(3) \over \pi^2} - {15 \zeta(5) \over \pi^4} \right) \,.
}
$F_S$ is exactly the $S^5$ free energy of a single real scalar field~\cite{Klebanov:2011gs}.  This then implies that
\es{F05}{
 F^{(1)}_{\rm min}  =  F_S \,.
}

One may perform similar calculations in the higher dimensional cases. In all even-dimensional spaces through AdS$_{12}$,  we have verified explicitly that $\zeta_\text{HS}'(0) = 0$ in the theory of all integer spins and $\zeta_\text{min~HS}'(0) = -2 F_S$ in the minimal theory. We expect this to be true in all even dimensional AdS spaces. In Appendix~\ref{AdS8} we describe the calculation in AdS$_8$ as an example.

\subsection{The interacting fixed point in $d=5$}
\label{5dfixed}
For $d>4$ we can construct interacting UV fixed points by perturbing free scalar theories with quartic operators.
The $U(N)$ singlet sector of $N$ complex scalars may be perturbed by the operator  $\frac {\lambda} {4} (\bar \phi^i\phi^i)^2$, while the
$O(N)$ singlet sector of $N$ real scalars -- by the operator
$\frac {\lambda} {4} (\phi^i\phi^i)^2$. Working in $4+\epsilon$ dimensions, it is not hard to see that the UV fixed point exists only when $\lambda < 0$, i.e.
the scalar potential is unstable \cite{Maldacena:2012sf} (this is related to such a theory being formally asymptotically free in $d=4$).
For example, for the $O(N)$ symmetric theory
\es{betafun}{
\beta_\lambda = \epsilon \lambda + \frac {N+8}{8 \pi^2} \lambda^2 + \ldots
}
so that the UV fixed point is at $\lambda_*=- \frac {8 \pi^2} {N+8}\epsilon + O(\epsilon^2)$.
However, for large $N$ one may hope that the UV theory is meta-stable.

In both the $U(N)$ symmetric and $O(N)$ symmetric cases, the dimension of the scalar operator at the UV fixed point is $\Delta_-= d-\Delta_+ +O(1/N)= 2 + O(1/N)$ \cite{Gubser:2002vv}.
For $d>6$ this value is below the unitary bound, while for $d=6$ it is right at the bound. However, it is unlikely that there exists an interacting $d=6$ large $N$ CFT
with a single-trace operator of dimension $\Delta=2$, because such a theory possesses a marginal triple-trace operator which is expected to have a non-vanishing
beta function (see, for example \cite{MuruganThesis}). Such a multi-trace instability would be a subtle effect from the point of view of the dual theory in AdS$_7$ with
the $\Delta=2$ scalar boundary condition.

It appears, therefore, that the only interesting case of an interacting unitary 
theory is $d=5$ \cite{Maldacena:2012sf}.
This interacting large $N$ UV fixed point should be dual to the Vasiliev theory in $AdS_6$ where the bulk scalar is quantized with the alternate, $\Delta=2$, boundary condition. In this case,
\es{spin05-alt}{
K_{(2,0)}  &=  {1 \over 60}  \int_{-1/2}^{0} dx \, x \left( x^2 - \frac14 \right) \left( x^2 - \frac94 \right) \psi\left( x + \frac12 \right) \\
&= K_{(3,0)} + {3 \zeta(5)  + \pi^2 \zeta(3) \over 48 \pi^4}  \,,
}
with $K_{(3,0)}$ given in~\eqref{spin05}. This implies
\es{Fuv}
{
F^{(1)}_{\rm UV} - F^{(1)}_{\rm IR} = - {3 \zeta(5)  + \pi^2 \zeta(3) \over 96 \pi^4} \approx -0.0016 \,,}
 which is equal to the results found in \cite{Diaz:2007an,Klebanov:2011gs,Giombi:2013yva}
 for the flow from the $\Delta=2$ to the $\Delta=3$ boundary condition in $d=5$.
The sign in (\ref{Fuv}) is consistent with the $5$ dimensional version of the $F$-theorem~\cite{Klebanov:2011gs}, since in that case it is conjectured that
$- F= \log Z_{S^5}$ decreases under RG flow.

\section*{Acknowledgments}

We thank A. Jevicki, J. Maldacena, M. Vasiliev, A. Zhiboedov, and especially A. Tseytlin, for helpful discussions. The work of SG was supported in part by the US NSF under Grant No.~PHY-1318681.
The work of IRK and BRS was supported in part by the US NSF under Grant No.~PHY-1314198.

 \appendix

 \section{Evaluating $\zeta_{(\Delta,s)}'(0)$ for odd $d$} \label{Method}

 In this Appendix we describe how to calculate $\zeta_{(\Delta,s)}'(0)$ in even-dimensional AdS spaces.   First, we use the identity $\tanh(\pi u) = 1 - 2 (1 + e^{2 \pi u})^{-1}$ to write
\es{zetaP0}{
\zeta'_{(\nu + d/2,s)}(0) = {g(s) \Gamma\left( - {d \over 2} \right) \over 2^d \sqrt{\pi} \Gamma \left( {d+1 \over 2} \right) } \big[ I_1(\nu) + I_2(\nu) \big] \,,
}
where
\es{I1I2}{
I_1(\nu) &= \lim_{z \to 0} {d \over d z} \left( \int_0^\infty d u \, u \left[ u^2 + \left( s + {d - 2 \over 2} \right)^2 \right] {\prod_{j =1/2}^{(d-4)/2} (u^2 + j^2)  \over (u^2 + \nu^2)^z } \right) \,, \\
I_2(\nu) &= 2 \int_0^\infty d u \, u \left[ u^2 + \left( s + {d - 2 \over 2} \right)^2 \right] {\prod_{j =1/2}^{(d-4)/2} (u^2 + j^2)  \log (u^2 + \nu^2) \over \big(1 + e^{2 \pi u}\big)} \,.
}
The integral $I_1(\nu)$ may be evaluated simply in a given dimension $d$ by first continuing to large $z$ so that the integral converges and then using the identity
\es{Useful}{
\lim_{z \to 0} {d \over d z} \left( \int_0^\infty d u \, {u^{2 p +1}  \over (u^2 + \nu^2)^z } \right) = (-1)^{p+1} \nu^{2(1 + p)} {H_{1 + p} - 2 \log \nu  \over 2(1 + p) } \,, \qquad p \in \mathbb{Z}_{\geq 0} \,.
}

The integral $I_2(\nu)$ is more subtle.  Below we describe a method for rewriting this integral in a more convenient form.  It is sufficient to consider
\es{fDeff}{
\int_0^\infty d u \, {u^{2 p +1} \log (u^2 + \nu^2)  \over \big(1 + e^{2 \pi u}\big)   } = 2 \,\int_0^\nu dx \, x \, A_p(x) +  \, c_p \,,
}
where we have defined
\es{Adeff}{
A_p(x)= \int_0^\infty d u {u^{2p+1} \over (u^2 + x^2) \big(1 + e^{2 \pi u} \big)} \,, \qquad  c_p = \int_0^\infty d u {u^{2p+1}  \log u^2 \over \big(1 + e^{2 \pi u} \big)} \,.
}
It was shown in~\cite{Camporesi:1991nw} that
\es{A0}{
A_0(x) = {1 \over 2} \psi \left( x + {1 \over 2} \right) - {1 \over 2} \log x \,,
}
and by straightforward manipulations we find the identity
\es{AIdent}{
A_{p+1}(x) = {(-1)^p \over 2^{2p+3} (p+1)}  \left( 2^{2p+1} - 1 \right) B\big(2p + 2 \big) - x^2 A_p(x) \,,
}
which allows us to calculate the $A_p(x)$ inductively.  Of course, we are still left with the integral over $x$ in~\eqref{fDeff}.
Following the procedure above leads to the known result~\eqref{d3Result} in $d = 3$~\cite{Camporesi:1993mz}.

\section{$\zeta_\text{HS}'(0)$ in AdS$_8 $} \label{AdS8}

In this Appendix, we outline the calculation of $\zeta_\text{HS}'(0)$ in AdS$_8$.  Using Appendix~\ref{Method}, we may write
\es{d5Resultzp}{
\zeta'_{(\Delta,s)}(0) = J_{(\Delta,s)} + K_{(\Delta,s)} \,,
}
with
\es{J7}{
J_{(\nu + 7/2,s)} &= {g(s) \over 2520} \left[ \nu^2 { 59845 + 8988 s (s+5) \over 161280} + \nu^4 {7413 + 1160 s (s + 5) \over 3840} \right. \\
&\left.- \nu^6 {313 + 36 s (s+5) \over 288}+ \nu^8 {11 \over 96}   + \frac{9}{40} \left( s +\frac52\right)^2 c_0 + \frac52 \left( s (s + 5) + \frac{259}{40} \right) c_1 \right. \\
&\left.+\left( s (s + 5) + \frac{35}{4} \right) c_2 + c_3 ] \right]
}
 and
\es{K7}{
K_{(\nu + 7/2,s)} &=- {g(s) \over 2520}  \int_0^\nu dx \, x \left( x^2 - \frac14 \right) \left( x^2 - \frac94 \right) \left[ x^2 - \left( s + {5 \over 2} \right)^2 \right] \psi\left( x + \frac12 \right) \,.
}
It is straightforward to verify that $J_{(\Delta,s)}$ does not contribute to $F^{(1)}$:
 \es{J72}{
 J_{(5,0)} + \sum_{s = 1}^\infty \big( J_{(5+s,s)} - J_{(6+s,s-1)} \big) = J_{(5,0)} + \sum_{s = 2, 4, \cdots}^\infty \big( J_{(5+s,s)} - J_{(6+s,s-1)} \big) =0 \,,
 }
 when the spin sums are regulated using~\eqref{altReg: 2}.

 Calculating the contribution from the $K_{(\Delta,s)}$ is less trivial.  The scalar integral evaluates simply to
 \es{K07}{
 K_{(5,0)} &= - {1 \over 2520}  \int_0^{3/2} dx \, x \left( x^2 - \frac14 \right) \left( x^2 - \frac94 \right) \left( x^2 - \frac{25}{4} \right) \psi\left( x + \frac12 \right) \\
 &= - {2171077 \over 722534400} - {15157 \log 2 \over 232243200} + {537 \log A \over 35840} - {\zeta(3) \over 160 \pi^2} - {\zeta(5) \over 64 \pi^2} + {3 \zeta(7) \over 128 \pi^6} \\
 &+ {13 \zeta'(-3) \over 3072} + {61 \zeta'(-5) \over 5120} + {17 \zeta'(-7) \over 21504}  \,.
}
Then, using the same regularization procedure described in the AdS$_4$ and AdS$_6$ sections, a long but straightforward calculation gives the expected results
\es{Ks7}{
\sum_{s = 1}^\infty \big( K_{(5+s,s)} - K_{(6+s,s-1)} \big) &= -  K_{(5,0)} \,,\\
\sum_{s = 2, 4, \dots}^\infty \big( K_{(5+s,s)} - K_{(6+s,s-1)} \big) &= -  K_{(5,0)} - 2 \, F_S \,,
}
where
\es{Fs7}{
F_S = {1 \over 2^{12}} \left( 4 \log 2 + {82 \zeta(3) \over 15 \pi^2} - {10 \zeta(5) \over \pi^4} - {63 \zeta(7) \over \pi^6} \right)
}
is the free energy of a real scalar field on $S^7$~\cite{Klebanov:2011gs}.

\bibliographystyle{ssg}
\bibliography{CGLP}

\end{document}